\documentclass[preprint,showpacs,preprintnumbers,amsmath,amssymb]{revtex4}
\usepackage{amssymb}
\usepackage{amsmath}
\usepackage{graphicx}
\usepackage{epsfig}
\usepackage{subfigure}
\usepackage{amsfonts}
\usepackage{CJK}

\begin{document}

\title{Crosstalk-insensitive method for simultaneously coupling
multiple pairs of resonators}

\author{Chui-Ping Yang$^{1,2,3}$, Qi-Ping Su$^{1,2,3}$, Shi-Biao Zheng$^{4}$, and Franco Nori$^{1,2}$}
\address{$^1$CEMS, RIKEN, Saitama 351-0198, Japan}
\address{$^2$Department of Physics, The University of Michigan, Ann Arbor, Michigan 48109-1040, USA}
\address{$^3$Department of Physics, Hangzhou Normal University, Hangzhou, Zhejiang 310036, China}
\address{$^4$Department of Physics, Fuzhou University, Fuzhou 350002, China}

\date{\today}

\begin{abstract} In a circuit consisting of two or more resonators, the inter-cavity
crosstalk is inevitable, which could create some problems, such as
degrading the performance of quantum operations and the fidelity of various
quantum states. The focus of this work is to propose a crosstalk-insensitive method
for simultaneously coupling multiple pairs of resonators, which is important
in large-scale quantum information processing and communication in a
network consisting of resonators or cavities. In this work, we consider $2N$ resonators
of different frequencies, which are coupled to a three-level quantum system (qutrit).~By
applying a strong pulse to the coupler qutrit, we show that an effective Hamiltonian can be constructed for
simultaneously coupling multiple pairs of resonators.~The main advantage of this proposal is that the effect of inter-resonator
crosstalks is greatly suppressed by using resonators of different frequencies.~In addition, by employing the qutrit-resonator
dispersive interaction, the intermediate higher-energy level of the
qutrit is virtually excited and thus decoherence from this level is
suppressed. This effective Hamiltonian can be applied to implement quantum operations with photonic qubits distributed
in different resonators. As one application of this Hamiltonian, we show how to simultaneously generate
multiple EPR pairs of photonic qubits distributed in $2N$ resonators. Numerical simulations show that it
is feasible to prepare two high-fidelity EPR photonic pairs using a setup of four one-dimensional transmission
line resonators coupled to a superconducting flux qutrit with current circuit QED technology.
\end{abstract}

\maketitle
\date{\today}

\begin{center}
\textbf{I. INTRODUCTION}
\end{center}

Circuit Quantum Electrodynamics (QED), consisting of microwave resonators
and superconducting qubits, has quickly developed in the last decade and is
considered one of the most promising platforms for quantum information
processing (QIP) (for reviews, see [1-4]). Superconducting qubits are very
important in solid-state quantum computation and QIP, due to the
controllability of their level spacings, the scalability of circuits, and
the improvement of coherence times [5-14]. High-quality-factor microwave
resonators have also drawn much attention because they have many
applications in QIP; for example, they can be used as quantum data buses
[15-18] and quantum memories [19,20]. A superconducting coplanar waveguide
resonator with a (loaded) quality factor $Q=10^6$ [21,22] or with an
internal quality factor above $10^7$ [23] was previously reported.
Superconducting microwave resonators with a loaded quality factor $%
Q=3.5\times 10^7$ have been recently demonstrated in experiments [24], for
which the single-photon lifetime can reach near one millisecond, while the
cavity mode remains strong-coupled with a superconducting qubit. The strong
or ultrastrong coupling between a superconducting qubit and a microwave
cavity has been experimentally observed [18,25-27]. Moreover, quantum
phenomena, such as squeezing or multiphoton quantum Rabi oscillations in the
ultrastrong coupling regimes, have been theoretically investigated [28,29].

As this is relevant to this work, here we provide a brief review on the
production and manipulation of quantum states of microwave photons in
circuit QED. For convenience, the term cavity and resonator will be used
interchangeably. During the past years, a number of theoretical works
[30-35] have been done on the preparation of Fock states, coherent states,
squeezed states, Schrodinger cat states and an arbitrary superposition of
Fock states of a single superconducting cavity. Experimentally, the creation
of a Fock state or a superposition of Fock states of a single
superconducting cavity has been reported [17,36,37]. In recent years,
attention has shifted to larger systems involving two or more cavities.
Based on circuit QED, many theoretical proposals have been presented for
implementing quantum state synthesis of photons in two resonators $[38,39$],
generating entangled photon Fock states of two resonators [40,41], creating
photon NOON states of two resonators [38,39,42,43], and preparing entangled
photon Fock states or entangled coherent states of more than two cavities
[44-46]. In addition, schemes for realizing two-qubit or multi-qubit quantum
gates with microwave photons distributed in different cavities have been
proposed [47,48]. Experimentally, the creation of photon NOON states of two
resonators has been reported [49], and the coherent transfer of microwave
photons between three resonators interconnected by two phase qubits has also
been demonstrated [50]. All these works are fundamental and important, and
open new avenues to use microwave photons as resource for quantum
computation and communication.

In a circuit consisting of two or more resonators, the inter-cavity
crosstalk is inevitable, which could create some problems, such as degrading
the performance of quantum operations and the fidelity of various quantum
states. Let us consider a two-cavity system, for which the inter-cavity
crosstalk is described by the Hamitonian $H=g(e^{i\Delta
t}ab^{+}+e^{-i\Delta t}a^{+}b)$, where $g$ is the inter-cavity crosstalk
strength between the two cavities, $\Delta $ is the detuning between the
frequencies of the two cavities, and $a$ ($b$) is the photon annihilation
operator of one (the other) cavity. From the form of the Hamiltonian $H$, it
can be seen that the effect of the cavity-cavity crosstalk depends on the
ratio of $\alpha =\Delta /g$, which increases as $\alpha $ decreases. In
other words, the effect of the cavity crosstalk is strongest for $\Delta =0$
(i.e., when the two cavities have the same frequency), while it can be
reduced by increasing $\alpha $ (e.g., increasing the detuning $\Delta $ for
a given $g$). The discussion here gives a hint on how to reduce the effect
of the inter-cavity crosstalk. Namely, in order to reduce the effect of the
inter-cavity crosstalk, one could employ cavities with different frequencies.

In this work, we focus on a physical system consisting of $2N$ resonators
coupled to a three-level quantum system (qutrit). It is shown that by
applying a strong pulse to the coupler qutrit, an effective Hamiltonian can
be obtained for simultaneously coupling multiple pairs of resonators with
different frequencies, which is insensitive to the inter-resonator
crosstalk. This effective Hamiltonian can be applied to implement quantum
operations with photonic qubits distributed in different resonators. The
major advantage of this proposal is that the inter-resonator crosstalk is
greatly reduced because of using different resonator frequencies. In
addition, the intermediate higher-energy level of the qutrit is virtually
excited due to the qutrit-resonator dispersive interaction, and thus
decoherence from this level is greatly suppressed.

As one application of this constructed Hamiltonian, we show how to
simultaneously generate multiple Einstein-Podolsky-Rosen (EPR) pairs [51] of
photonic qubits distributed in $2N$ resonators. The prepared EPR pairs are
particularly useful in quantum communication and QIP. As a specific
experimental realization, we further discuss the possible experimental
implementation of two EPR pairs of photonic qubits using a setup consisting
of four one-dimensional transmission line resonators coupled to a
superconducting flux qutrit. With realistic device and circuit parameters,
numerical simulations show that the fidelity can reach $98.42\%$ for the
joint state of two EPR pairs and is no less than $99.05\%$ for each EPR pair.

We note that previous works focused on the preparation of a \textit{single}
EPR pair in various physical systems, such as neutral Kaons [52], trapped
ions [53,54], atoms interacting with a cavity mode [55-58], Bose-Einstein
condensates [59-61], two harmonic oscillators in nonequilibrium open systems
[62], center-of-mass motion of two massive objects [63], and superconducting
qubits [15,64-66]. In stark contrast, ours is aimed at simultaneously
generating \textit{multiple} EPR pairs by using the constructed Hamiltonian,
which is insensitive to the inter-resonator crosstalk.

This paper is organized as follows. In Sec.~II, we derive the effective
Hamiltonian governing the dynamics of $N$ pairs of cavities plus one
three-level coupler qutrit. This Hamiltonian describes paired interactions
between these cavities in parallel without interfering with each other. In
Sec.~III, we show in detail how to simultaneously prepare $N$ pairs of
photonic qubits using this effective Hamiltonian. In Sec.~IV, we discuss the
possible experimental implementation of generating EPR states of two pairs
of photonic qubits in circuit QED. In Sec.~V, we summarize our results and
discuss other possible applications of this physical process.

\begin{center}
\textbf{II. EFFECTIVE HAMILTONIAN}
\end{center}

\begin{figure}[tbp]
\begin{center}
\includegraphics[bb=186 445 418 607, width=8.5 cm, clip]{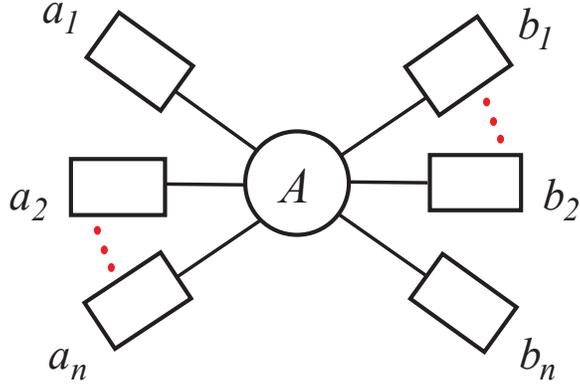} \vspace*{%
-0.08in}
\end{center}
\caption{(color online) Diagram of a coupler qutrit $A$ (the circle at the
center) and $2n$ coupled resonators. Each rectangle represents a resonator.
The coupler qutrit $A$ can be an artificial atom, such as a quantum dot or a
superconducting qutrit capacitively/inductively coupled to each resonator.}
\label{fig:1}
\end{figure}

\begin{figure}[tbp]
\begin{center}
\includegraphics[bb=121 288 450 624, width=8.5 cm, clip]{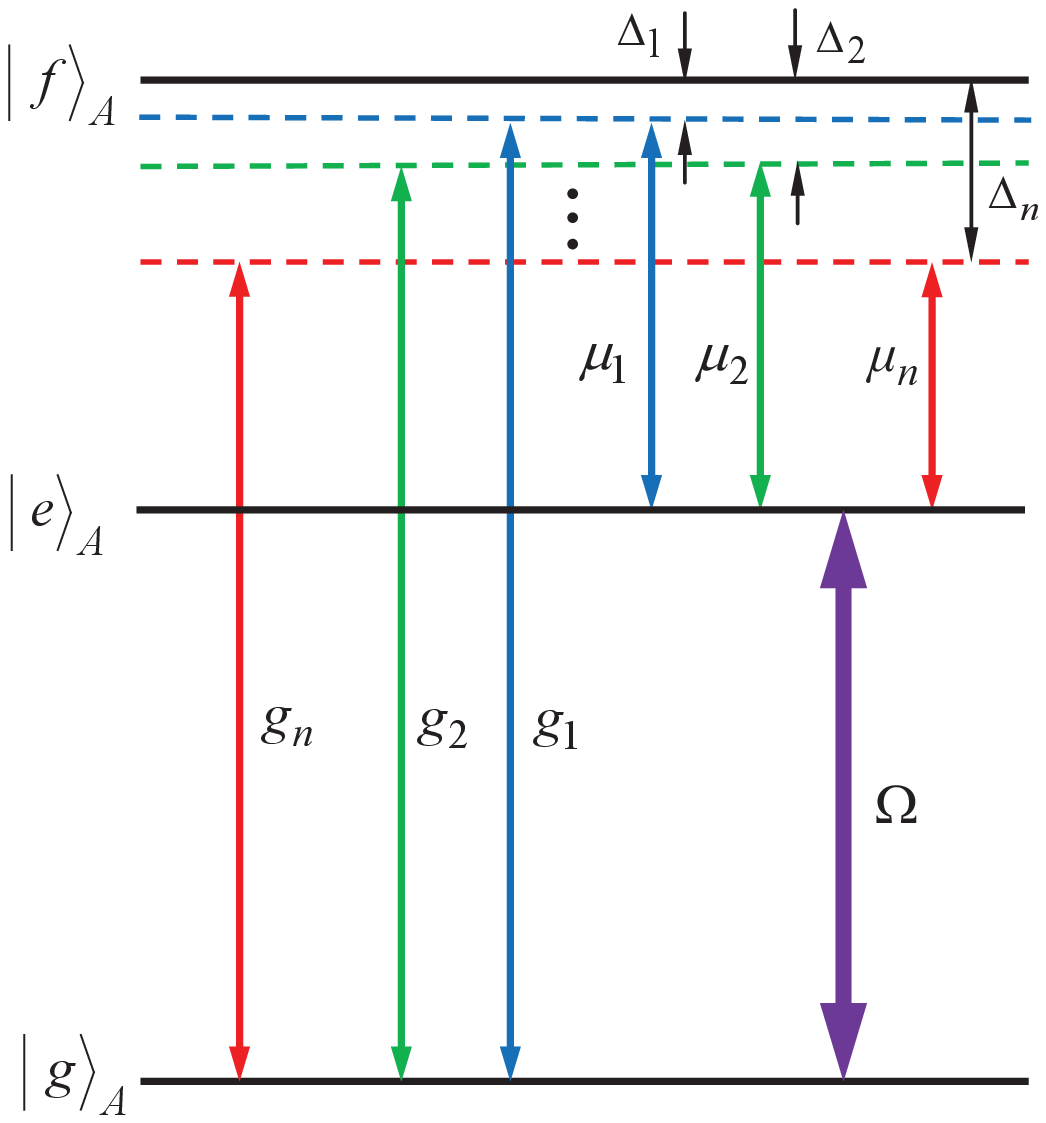} \vspace*{%
-0.08in}
\end{center}
\caption{(Color online) Illustration of the qutrit-resonator dispersive
interaction. The $\left\vert g\right\rangle $ $\leftrightarrow $ $\left\vert
f\right\rangle $ transition of the qutrit is simultaneously coupled to the $%
n $ resonators ($a_1,a_2,...,a_n$), with coupling constants $g_1,g_2,...,g_n$
and detunings $\Delta_1,\Delta_2,...,\Delta_n$, respectively. The $%
\left\vert e\right\rangle $ $\leftrightarrow$ $\left\vert f\right\rangle $
transition of the qutrit is simultaneously coupled to the other $n$
resonators ($b_1,b_2,...,b_n$), with coupling constants $\protect\mu_1,%
\protect\mu_2,...,\protect\mu_n$ and detunings $\Delta_1,\Delta_2,...,%
\Delta_n$, respectively. In addition, a microwave pulse is resonantly
coupled to the $\left\vert g\right\rangle $ $\leftrightarrow $ $\left\vert
e\right\rangle $ transition of the qutrit, with a Rabi frequency $\Omega$.}
\label{fig:2}
\end{figure}

Consider $2N$ resonators coupled to a qutrit $A$ (Fig.~1). The first set of $%
N$ resonators are labeled as resonators $a_{1},a_{2},...,$ and $a_{N}$ while
the second set of $N$ resonators are labeled as resonators $b_{1},b_{2},...,$%
and $b_{N}$. In addition, the three levels of qutrit $A$ are denoted as $%
\left\vert g\right\rangle ,$ $\left\vert e\right\rangle $ and $\left\vert
f\right\rangle $ (Fig.~2). Suppose that resonator $a_{j}$ ($b_{j}$) with $%
j=1,2,...,N$ is coupled to the $\left\vert g\right\rangle $ $\leftrightarrow
$ $\left\vert f\right\rangle $ $\left( \left\vert e\right\rangle
\leftrightarrow \left\vert f\right\rangle \right) $ transition with coupling
strength $g_{j}$ ($\mu _{j}$) and detuning $\Delta _{j}=\omega _{fg}-\omega
_{a_{j}}=\omega _{fe}-\omega _{b_{j}}>0$ (Fig.~2)$.$ Here, $\omega _{a_{j}}$
($\omega _{b_{j}}$) is the frequency of resonator $a_{j}$ ($b_{j}$). In
addition, a classical pulse of frequency $\omega $ is applied to the qutrit $%
A,$ which is resonant with the $\left\vert g\right\rangle $ $\leftrightarrow
$ $\left\vert e\right\rangle $ transition (Fig.~2)$.$ In the interaction
picture, the Hamiltonian of the whole system is given by (assuming $\hbar =1$%
)
\begin{eqnarray}
H &=&\sum_{j=1}^{N}\left( g_{j}e^{i\Delta _{j}t}\hat{a}_{j}S_{fg}^{+}+\text{%
H.c.}\right) +\sum_{j=1}^{N}\left( \mu _{j}e^{i\Delta _{j}t}\hat{b}%
_{j}S_{fe}^{+}+\text{H.c.}\right)  \notag \\
&&\ +\left( \Omega S_{eg}^{+}+\text{H.c.}\right) ,
\end{eqnarray}
where $S_{fg}^{+}=\left\vert f\right\rangle \left\langle g\right\vert $, $%
S_{fe}^{+}=\left\vert f\right\rangle \left\langle e\right\vert ,$ $%
S_{eg}^{+}=\left\vert e\right\rangle \left\langle g\right\vert ,$ $\Omega $
is the Rabi frequency of the pulse, and $\hat{a}_{j}$ ($\hat{b}_{j}$) is the
photon annihilation operator of resonator $a_{j}$ ($b_{j}$).

Under the large-detuning condition $\Delta _{j}\gg g_{j},\mu _{j},$ the
intermediate level $\left\vert f\right\rangle $ can be adiabatically
eliminated, and the Raman transitions between the states $\left\vert
g\right\rangle $ and $\left\vert e\right\rangle $ are induced by resonator
pairs $\left( a_{j},b_{j}\right) $ ($j=1,2,...,N$). Under the following
condition
\begin{equation}
\frac{\left\vert \Delta _{j}-\Delta _{k}\right\vert }{\Delta
_{j}^{-1}+\Delta _{k}^{-1}}\gg g_{j}g_{k},\;g_{j}\mu _{k},\;\mu _{j}\mu
_{k};\;\text{ }j\neq k,
\end{equation}%
the Raman couplings associated with resonator pairs $\left(
a_{j},a_{k}\right) ,$ $(b_{j},b_{k}),$ and $\left( a_{j},b_{k}\right) $,
with $j\neq k$, are suppressed because the corresponding effective coupling
strengths are much smaller than the detunings of these Raman transitions. In
addition, we assume $\Delta _{j}\gg \Omega $ so that the energy shift of the
qutrit dressed states produced by the classical pulse is very small compared
to $\Delta _{j}$, and hence the effect of this pulse on the strength of each
Raman coupling is negligible. Under these conditions, we can obtain the
following effective Hamiltonian [67,68]

\begin{eqnarray}
H_{\mathrm{eff}} &=&-\sum_{j=1}^{N}\frac{g_{j}^{2}}{\Delta _{j}}\hat{a}_{j}%
\hat{a}_{j}^{+}\left\vert g\right\rangle \left\langle g\right\vert
-\sum_{j=1}^{N}\frac{\mu _{j}^{2}}{\Delta _{j}}\hat{b}_{j}\hat{b}%
_{j}^{+}\left\vert e\right\rangle \left\langle e\right\vert  \notag \\
&&\ -\sum_{j=1}^{N}\lambda _{j}(\hat{a}_{j}\hat{b}_{j}^{+}S_{eg}^{+}+\hat{a}%
_{j}^{+}\hat{b}_{j}S_{eg}^{-})  \notag \\
&&\ +\Omega S_{x},
\end{eqnarray}
where $S_{eg}^{-}=\left\vert g\right\rangle \left\langle e\right\vert ,$ $%
S_{x}=S_{eg}^{+}+S_{eg}^{-},$\ and $\lambda _{j}=\frac{g_{j}\mu _{j}}{\Delta
_{j}}.$ Here, the terms in the first line are ac-Stark shifts of the level $%
\left\vert g\right\rangle $\ ($\left\vert e\right\rangle $) induced by the
resonator mode $a_{j}$\ ($b_{j}$). The terms in the second line represent
the Raman couplings induced by the $N$ pairs of cavities.

In a rotated basis \{$|+\rangle ,|-\rangle $\} with $|\pm \rangle
=(|g\rangle \pm |e\rangle )/\sqrt{2}$, one has $S_{eg}^{+}=\left( \widetilde{%
S}_{z}+\widetilde{S}^{+}-\widetilde{S}^{-}\right) /2,$ $S_{eg}^{-}=\left(
\widetilde{S}_{z}-\widetilde{S}^{+}+\widetilde{S}^{-}\right) /2$, and $S_{x}=%
\widetilde{S}_{z}$, where $\widetilde{S}_{z}=|+\rangle \langle +|-|-\rangle
\langle -|$ , $\widetilde{S}^{+}=|+\rangle \langle -|,$ and $\widetilde{S}%
^{-}=|-\rangle \langle +|$. In addition, one has $\left\vert g\right\rangle
\left\langle g\right\vert =\frac{1}{2}\left( I+\widetilde{S}^{+}+\widetilde{S%
}^{-}\right) $ and $\left\vert e\right\rangle \left\langle e\right\vert =%
\frac{1}{2}\left( I-\widetilde{S}^{+}-\widetilde{S}^{-}\right) .$

Performing the unitary transformation $e^{iH_{0}t}$, with
\begin{equation}
H_{0}=\Omega S_{x}=\Omega \widetilde{S}_{z},
\end{equation}
one obtains
\begin{eqnarray}
\widetilde{H}_{\mathrm{eff}} &=&e^{iH_{0}t}(H_{\mathrm{eff}%
}-H_{0})e^{-iH_{0}t}  \notag \\
\ &=&-\frac{1}{2}\left( \sum_{j=1}^{N}\frac{g_{j}^{2}}{\Delta _{j}}\hat{a}%
_{j}\hat{a}_{j}^{+}\right) \left( I+e^{i2\Omega t}\widetilde{S}%
^{+}+e^{-i2\Omega t}\widetilde{S}^{-}\right)  \notag \\
&&\ \ -\frac{1}{2}\left( \sum_{j=1}^{N}\frac{\mu _{j}^{2}}{\Delta _{j}}\hat{b%
}_{j}\hat{b}_{j}^{+}\right) \left( I-e^{i2\Omega t}\widetilde{S}%
^{+}-e^{-i2\Omega t}\widetilde{S}^{-}\right)  \notag \\
&&\ \ \ -\sum_{j=1}^{n}\frac{\lambda _{j}}{2}\left[ \hat{a}_{j}\hat{b}%
_{j}^{+}(\widetilde{S}_{z}+e^{i2\Omega t}\widetilde{S}^{+}-e^{-i2\Omega t}%
\widetilde{S}^{-})\right.  \notag \\
&&\ \ \left. +\hat{a}_{j}^{+}\hat{b}_{j}(\widetilde{S}_{z}-e^{i2\Omega t}%
\widetilde{S}^{+}+e^{-i2\Omega t}\widetilde{S}^{-})\right] .
\end{eqnarray}
In the strong driving regime $\Omega \gg \frac{g_{j}^{2}}{4\Delta _{j}},%
\frac{\mu _{j}^{2}}{4\Delta _{j}},\frac{\lambda _{j}}{4}$, one can apply a
rotating-wave approximation and discard the terms that oscillate with high
frequencies. Thus, the above Hamiltonian reduces to
\begin{eqnarray}
\widetilde{H}_{\mathrm{eff}} &=&-\frac{1}{2}\sum_{j=1}^{N}\left( \frac{%
g_{j}^{2}}{\Delta _{j}}\hat{a}_{j}\hat{a}_{j}^{+}+\frac{\mu _{j}^{2}}{\Delta
_{j}}\hat{b}_{j}\hat{b}_{j}^{+}\right) \otimes I  \notag \\
&&\ \ \ \ -\sum_{j=1}^{N}\frac{\lambda _{j}}{2}\left( \hat{a}_{j}\hat{b}%
_{j}^{+}+\hat{a}_{j}^{+}\hat{b}_{j}\right) \widetilde{S}_{z}.
\end{eqnarray}

Performing the additional unitary transformation $e^{iH_{0}^{\prime }t}$,
with
\begin{equation}
H_{0}^{\prime }=-\frac{1}{2}\sum_{j=1}^{N}\left( \frac{g_{j}^{2}}{\Delta _{j}%
}\hat{a}_{j}\hat{a}_{j}^{+}+\frac{\mu _{j}^{2}}{\Delta _{j}}\hat{b}_{j}\hat{b%
}_{j}^{+}\right) \otimes I,
\end{equation}%
we have
\begin{eqnarray}
H_{e} &=&e^{iH_{0}^{\prime }t}\left( \widetilde{H}_{\mathrm{eff}%
}-H_{0}^{^{\prime }}\right) e^{-iH_{0}^{\prime }t}  \notag \\
\ \ &=&\ -\sum_{j=1}^{N}\frac{\lambda _{j}}{2}\left( e^{i\delta _{j}t}\hat{a}%
_{j}\hat{b}_{j}^{+}+e^{-i\delta _{j}t}\hat{a}_{j}^{+}\hat{b}_{j}\right)
\widetilde{S}_{z},
\end{eqnarray}%
where $\delta _{j}=\frac{g_{j}^{2}-\mu _{j}^{2}}{\Delta _{j}}.$ In the
following, we set $g_{j}=\mu _{j}$ (achievable by tuning the coupling
capacitance between the qubit and resonator $a_{j}$, as well as the coupling
capacitance between the qubit and resonator $b_{j}$), resulting in
\begin{equation}
H_{e}=-\sum_{j=1}^{N}\frac{\lambda _{j}}{2}\left( \hat{a}_{j}\hat{b}_{j}^{+}+%
\hat{a}_{j}^{+}\hat{b}_{j}\right) \widetilde{S}_{z}.
\end{equation}

We note that previous works [69,70] considered the coupling of a quantized
cavity mode and a strong classical pulse via a superconducting qubit or
two-level atoms, which also employed the strong driving limit in the
derivation of the effective Hamiltonians. In this sense, they are related to
this work. However, they [69,70] are different from ours. The reasons are:
they were focused on how to construct the simultaneous implementation of a
Jaynes-Cummings and anti-Jaynes-Cummings dynamics for a system composed of a
superconducting qubit/two-level atoms and \textit{one} cavity, and only a
single quantized cavity mode was involved there. Instead, our work is aimed
at deriving an effective Hamiltonian for simultaneously coupling \textit{%
multiple} pairs of resonators. One can see that the form of our effective
Hamiltonian of Eq. (9) is different from those given in [69,70]. In
addition, our effective Hamiltonian (9) contains two quantized cavity modes
for each pair of resonators, instead of just one single cavity mode.

\begin{center}
\textbf{III. GENERATION OF MULTIPLE EPR STATES}
\end{center}

When the qutrit is in the state $\left| +\right\rangle $ (readily prepared
by applying a $\pi $-pulse resonant with the $\left| g\right\rangle
\leftrightarrow \left| e\right\rangle $ transition of the qutrit initially
in the ground state $\left| g\right\rangle $), it will remain in this state
because the state $\left| +\right\rangle $ is not affected by the
Hamiltonian (9). Thus, the qutrit part can be ignored and the effective
Hamiltonian (9) further reduces to
\begin{equation}
H_{e}=-\sum_{j=1}^{N}\frac{\lambda _{j}}{2}(\hat{a}_{j}\hat{b}_{j}^{+}+\hat{a%
}_{j}^{+}\hat{b}_{j}).
\end{equation}
This Hamiltonian describes the coupler-mediated effective interactions for
the $N$ pairs of cavities $\left( a_{j},b_{j}\right) $ in parallel, which
will be used below to simultaneously generate multiple EPR states of $N$
pairs of photonic qubits.

Note that the Hamiltonian (10) is obtained under unitary transformations $%
e^{iH_0t}$ and $e^{iH_0^{\prime }t}$. To obtain the time-propagating states
in the original internation picture, two reverse transformations $e^{-iH_0t}$
and $e^{-iH_0^{\prime }t}$ need to be performed on the corresponding
time-evolution states under this Hamiltonian.

Assume now that the first set of resonators $(a_{1},a_{2},...,a_{N})$ is
initially in the state $\left\vert \psi \left( 0\right) \right\rangle
_{a}=\prod_{j=1}^{N}\left\vert 1\right\rangle _{a_{j}},$ i.e., each
resonator in this set is initially prepared in the single-photon state;
while the second set of resonators $(b_{1},b_{2},...,b_{N})$ is initially in
the state $\left\vert \psi \left( 0\right) \right\rangle _{b}=$\emph{\ }$%
\prod_{j=1}^{N}\left\vert 0\right\rangle _{b_{j}}$, i.e., each of these
resonators is initially prepared in the vacuum state.

One can easily find that under the Hamiltonian $H_{e},$ the state of the
resonator system after an evolution time $t$ is given by
\begin{eqnarray}
\left\vert \psi \left( t\right) \right\rangle _{ab}
&=&e^{-iH_{e}t}\left\vert \psi \left( 0\right) \right\rangle _{a}\left\vert
\psi \left( 0\right) \right\rangle _{b}  \notag \\
\ &=&e^{-iH_{e}t}\prod_{j=1}^{N}\left( \left\vert 1\right\rangle
_{a_{j}}\left\vert 0\right\rangle _{b_{j}}\right)  \notag \\
\ &=&e^{-iH_{e}t}\prod_{j=1}^{N}\left( a_{j}^{+}\left\vert 0\right\rangle
_{a_{j}}\left\vert 0\right\rangle _{b_{j}}\right)  \notag \\
\ &=&\prod_{j=1}^{N}\left[ \left( e^{-iH_{e}t}a_{j}^{+}e^{iH_{e}t}\right)
e^{-iH_{e}t}\left\vert 0\right\rangle _{a_{j}}\left\vert 0\right\rangle
_{b_{j}}\right]  \notag \\
\ &=&\prod_{j=1}^{N}\left\{ \left[ \cos \left( \frac{\lambda _{j}t}{2}%
\right) \hat{a}_{j}^{+}+i\sin \left( \frac{\lambda _{j}t}{2}\right) \hat{b}%
_{j}^{+}\right] \left\vert 0\right\rangle _{a_{j}}\left\vert 0\right\rangle
_{b_{j}}\right\}  \notag \\
\ &=&\prod_{j=1}^{N}\left[ \cos \left( \frac{\lambda _{j}t}{2}\right)
\left\vert 1\right\rangle _{a_{j}}\left\vert 0\right\rangle _{b_{j}}+i\sin
\left( \frac{\lambda _{j}t}{2}\right) \left\vert 0\right\rangle
_{a_{j}}\left\vert 1\right\rangle _{b_{j}}\right] \ ,\
\end{eqnarray}
where we have used the relation $e^{-iH_{e}t}\left\vert 0\right\rangle
_{a_{j}}\left\vert 0\right\rangle _{b_{j}}=\left\vert 0\right\rangle
_{a_{j}}\left\vert 0\right\rangle _{b_{j}}$. After returning to the original
interaction picture, the state of the whole system is given by
\begin{eqnarray}
\left\vert \Phi \left( t\right) \right\rangle _{abA}^{\prime }
&=&e^{-iH_{0}t}e^{-iH_{0}^{\prime }t}\left\vert \psi \left( t\right)
\right\rangle _{ab}\left\vert \varphi \left( t\right) \right\rangle _{A}
\notag \\
\ &=&\left\vert \psi \left( t\right) \right\rangle _{ab}^{\prime }\otimes
\left\vert \varphi \left( t\right) \right\rangle _{A},\;\;
\end{eqnarray}
where a common phase factor $e^{-i\Omega t}$ is dropped. Here, $\left\vert
\varphi \left( t\right) \right\rangle _{A}=\left\vert \varphi \left(
0\right) \right\rangle _{A}=\left\vert +\right\rangle ,$ and
\begin{eqnarray}
\left\vert \psi \left( t\right) \right\rangle _{ab}^{\prime }
&=&\prod_{j=1}^{N}\left[ e^{ig_{j}^{2}t/\Delta _{j}}e^{i\mu
_{j}^{2}t/(2\Delta _{j})}\cos \left( \frac{\lambda _{j}t}{2}\right)
\left\vert 1\right\rangle _{a_{j}}\left\vert 0\right\rangle
_{b_{j}}+ie^{ig_{j}^{2}t/(2\Delta _{j})}e^{i\mu _{j}^{2}t/\Delta _{j}}\sin
\left( \frac{\lambda _{j}t}{2}\right) \left\vert 0\right\rangle
_{a_{j}}\left\vert 1\right\rangle _{b_{j}}\right]  \notag \\
\ &=&e^{i3n\lambda t/2}\prod_{j=1}^{N}\left[ \cos \left( \frac{\lambda t}{2}%
\right) \left\vert 1\right\rangle _{a_{j}}\left\vert 0\right\rangle
_{b_{j}}+i\sin \left( \frac{\lambda t}{2}\right) \left\vert 0\right\rangle
_{a_{j}}\left\vert 1\right\rangle _{b_{j}}\right] ,
\end{eqnarray}
where we have used $g_{j}=\mu _{j}$ (set above) and $g_{j}^{2}/\Delta
_{j}=\lambda _{j}\equiv \lambda $ (independing of $j$). It can be seen from
Eq.~(13) that for $t=\pi /\left( 2\lambda \right) ,$ the $2N$ resonators are
prepared in the following state
\begin{equation}
\prod_{j=1}^{N}\left\vert \text{EPR}\right\rangle
_{a_{j}b_{j}}=\prod_{j=1}^{N}\frac{1}{\sqrt{2}}\left( \left\vert
1\right\rangle _{a_{j}}\left\vert 0\right\rangle _{b_{j}}+i\left\vert
0\right\rangle _{a_{j}}\left\vert 1\right\rangle _{b_{j}}\right) ,
\end{equation}
which is a product of $N$ EPR pairs of photonic qubits. This result implies
that the $N$ EPR photonic pairs are simultaneously generated after the
operation. Here, $\left\vert 0\right\rangle _{a_{j}}$ ($\left\vert
0\right\rangle _{b_{j}}$) and $\left\vert 1\right\rangle _{a_{j}}$ ($%
\left\vert 1\right\rangle _{b_{j}}$) represent the two logic states of the
photonic qubit $a_{j}$($b_{j}$).

Note that the above-mentioned condition $g_{j}^{2}/\Delta _{j}=\lambda
_{j}\equiv \lambda $ can be rewritten as
\begin{equation}
g_{j}^{2}/\Delta _{j}=g_{k}^{2}/\Delta _{k},\text{ }(j\neq k),
\end{equation}%
which can be readily met by adjusting the detuning $\Delta _{j}$ or $\Delta
_{k}$ (e.g., varying the resonator frequency). Alternatively, this condition
can be satisfied by adjusting the coupling strength $g_{j}$ or $g_{k}$
(e..g, through a prior design of the sample with appropriate
qutrit-resonator coupling capacitances).

As shown above, the $N$ EPR photonic pairs are prepared based on the
effective Hamiltonian (9), which was derived without considering the
unwanted couplings of the resonators/pulse with the irrelevant level
transitions of the qutrit. To minimize decoherence effects induced due to
the leakage into the level $\left\vert f\right\rangle $, one can employ the
DRAG pulse with a cosine envelope shape [71] or a detuned pulse with the
DRAG pulse shaping [72], which can significantly reduce both leakage error
and phase error. In addition, one can design the qutrit level structure with
a large level anharmonicity, such that the unwanted couplings of the
resonators/pulse with the irrelevant qutrit level transitions are
negligible. The strong pulse may cause heating of the qutrit. For a short
pulse, the effect of incoherent processes caused due to the heating, such as
thermal excitations or noise at the $\left\vert e\right\rangle
\leftrightarrow \left\vert f\right\rangle $ transition, is negligible [72].
For a long pulse, the unwanted incoherent processes induced by the heating
can be suppressed through improved thermalization by cooling the sample [73].

\begin{center}
\textbf{IV. POSSIBLE EXPERIMENTAL IMPLEMENTATION}
\end{center}

\begin{figure}[tbp]
\begin{center}
\includegraphics[bb=0 0 598 367, width=8.5 cm, clip]{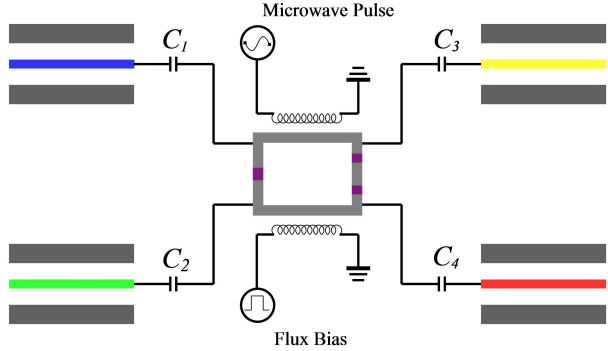} \vspace*{%
-0.08in}
\end{center}
\caption{(color online). Diagram of a setup for four one-dimensional
transmission line resonators coupled to a superconducting flux qutrit via
capacitances $C_1, C_2, C_3,$ and $C_4$, respectively.}
\label{fig:3}
\end{figure}

We now provide a quantitative analysis on the experimental feasibility of
the proposal. As an example, let us consider a setup of four one-dimensional
transmission line resonators coupled by a superconducting flux qutrit (Fig.
3).

With the unwanted interaction being considered, the Hamiltonian (1) is
modified as $H^{\prime }=H+\delta\!H_{1}+\delta\!H_{2}$ (with $N=2$), where $%
\delta\!H_{1}$ describes the unwanted inter-resonator crosstalk while $%
\delta\!H_{2}$ describes the unwanted $\left\vert e\right\rangle
\leftrightarrow \left\vert f\right\rangle $ transition induced by the pulse.
The expression of $\delta\!H_{1}$ is given by
\begin{eqnarray}
\delta\!H_{1} &=&g_{a_{1}b_{1}}e^{i\Delta
_{a_{1}b_{1}}t}a_{1}b_{1}^{+}+g_{a_{1}b_{2}}e^{i\Delta
_{a_{1}b_{2}}t}a_{1}b_{2}^{+}  \notag \\
&&+g_{a_{2}b_{1}}e^{i\Delta
_{a_{2}b_{1}}t}a_{2}b_{1}^{+}+g_{a_{2}b_{2}}e^{i\Delta
_{a_{2}b_{2}}t}a_{2}b_{2}^{+}  \notag \\
&&+g_{a_{1}a_{2}}e^{i\Delta
_{a_{1}a_{2}}t}a_{1}a_{2}^{+}+g_{b_{1}b_{2}}e^{i\Delta
_{b_{1}b_{2}}t}b_{1}b_{2}^{+}+h.c.,
\end{eqnarray}
where $g_{a_{j}b_{k}}$ is the coupling strength between the two resonators $%
a_{j}$ and $b_{k}$ with\textbf{\ }frequency detuning $\Delta
_{a_{j}b_{k}}=\omega _{b_{k}}-\omega _{a_{j}}$ ($j,k=1,2$)$;$ $%
g_{a_{1}a_{2}} $ is the coupling strength between the two resonators $a_{1}$
and $a_{2}$ with frequency detuning $\Delta _{a_{1}a_{2}}=\omega
_{a_{2}}-\omega _{a_{1}};$ and $g_{b_{1}b_{2}}$ is the coupling strength
between the two resonators $b_{1}$ and $b_{2}$ with frequency detuning $%
\Delta _{b_{1}b_{2}}=\omega _{b_{2}}-\omega _{b_{1}}$. $\delta\!H_{2}$ is
given by
\begin{equation}
\delta\!H_{2}=\Omega _{fe}e^{i\Delta t}S_{fe}^{+}+\text{H.c.},
\end{equation}
where $\Delta =\omega _{fe}-\omega _{eg},$ and $\Omega _{fe}$ is the pulse
Rabi frequency associated with the $\left\vert e\right\rangle
\leftrightarrow \left\vert f\right\rangle $ transition.

It should be mentioned that the $\left| g\right\rangle \leftrightarrow
\left| f\right\rangle $ transition induced by the pulse is negligible
because $\omega _{eg}\ll\omega _{fg}$ (Fig.~2). For simplicity, we also
assume that the resonator-induced coherent transitions between any other
irrelevant levels are negligibly small. This can be achieved by a prior
design of the coupler with a sufficiently large anharmonicity of the level
spacings (readily available for a superconducting flux device).

Taking into account the qutrit dephasing and energy relaxation as well as
the resonator dissipation, the system dynamics, under the Markovian
approximation, is determined by the master equation
\begin{eqnarray}
\frac{d\rho }{dt} &=&-i\left[ H^{\prime },\rho \right] +\sum%
\limits_{j=1}^{2}\kappa _{a_{j}}\mathcal{L}\left[ a_{j}\right]
+\sum\limits_{j=1}^{2}\kappa _{b_{j}}\mathcal{L}\left[ b_{j}\right]   \notag
\\
&&+\gamma _{fe}\mathcal{L}\left[ \sigma _{fe}^{-}\right] +\gamma _{fg}%
\mathcal{L}\left[ \sigma _{fg}^{-}\right] +\gamma _{eg}\mathcal{L}\left[
\sigma _{eg}^{-}\right]   \notag \\
&&+\sum\limits_{l=e,f}\gamma _{\varphi ,l}\left( \sigma _{ll}\rho \sigma
_{ll}-\sigma _{ll}\rho /2-\rho \sigma _{ll}/2\right) ,
\end{eqnarray}%
where $\mathcal{L}\left[ \Lambda \right] =\Lambda \rho \Lambda ^{+}-\Lambda
^{+}\Lambda \rho /2-\rho \Lambda ^{+}\Lambda /2$ (with $\Lambda
=a_{j},b_{j},\sigma _{fe}^{-},\sigma _{fg}^{-},\sigma _{eg}^{-})$,\ $\sigma
_{ee}=\left\vert e\right\rangle \left\langle e\right\vert ,$ and $\sigma
_{ff}=\left\vert f\right\rangle \left\langle f\right\vert $. In addition, $%
\kappa _{a_{j}}$ ($k_{b_{j}}$) is the decay rate of resonator $a_{j}$ ($b_{j}
$);\ $\gamma _{eg}$\ is the energy relaxation rate for the level $\left\vert
e\right\rangle $\ associated with the decay path $\left\vert e\right\rangle
\rightarrow \left\vert g\right\rangle $; $\gamma _{fe}$\ ($\gamma _{fg}$) is
the relaxation rate for the level $\left\vert f\right\rangle $ related to
the decay path $\left\vert f\right\rangle \rightarrow \left\vert
e\right\rangle $ ($\left\vert f\right\rangle \rightarrow \left\vert
g\right\rangle $); $\gamma _{\varphi ,e}$ ($\gamma _{\varphi ,f}$) is the
dephasing rate of the level $\left\vert e\right\rangle $ ($\left\vert
f\right\rangle $)\textbf{. }For numerical calculations, here we use the
QuTiP software [74,75]. QuTiP is an open-source software for simulating the
dynamics of open quantum systems, which can transfer quantum objects to
matrices and solve master equations numerically by using an ordinary
differential equation solver.

\begin{figure}[tbp]
\begin{center}
\includegraphics[bb=0 0 759 481, width=8.5 cm, clip]{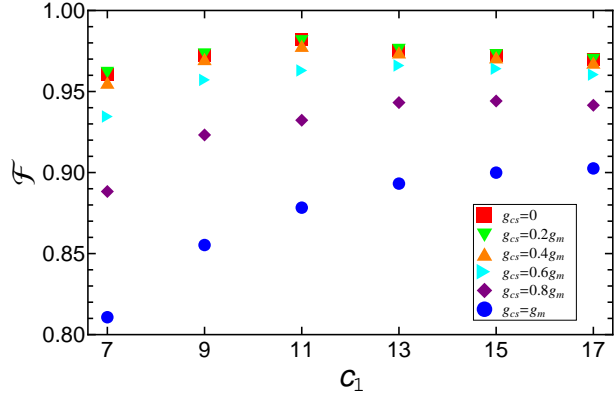} \vspace*{%
-0.08in}
\end{center}
\caption{(Color online) Fidelity versus the normalized detuning $c_1=\Delta
_{1}/g_{1}$. Here and in Fig. 5, there is a relation, $c_2=\Delta_2/g_2=%
\protect\sqrt{2}c_{1}$, for the $\Delta_1$ and $\Delta_2$ used in the
numerical simulations.}
\label{fig:4}
\end{figure}

The fidelity of the prepared two EPR states for the two pairs of photonic
qubits is given by $\mathcal{F}=\sqrt{\left\langle \psi _{\mathrm{id}%
}\right| \widetilde{\rho }\left| \psi _{\mathrm{id}}\right\rangle }.$ Here, $%
\left| \psi _{\mathrm{id}}\right\rangle =\left| \text{EPR}\right\rangle
_{a_1b_1}\otimes \left| \text{EPR}\right\rangle _{a_2b_2}$ is for the ideal
case; while $\widetilde{\rho }$ is the reduced density operator of the two
pairs of photonic qubits after tracing $\rho $ over the degrees of the
coupler qutrit, when the operation is performed in a realistic system (with
dissipation and dephasing considered).

In a real situation, it may be a challenge to obtain homogeneous coupling
strengths. Thus, we consider $\mu _1=0.95g_1$ and $\mu _2=0.95g_2$ in our
numerical simulation. Namely, there exists a difference of $5\%$ between the
coupling strengths for each pair of resonators, which may be reasonable in
experiments. Note that our numerical simulations are performed by choosing
the operation time $t=\pi /\left( 2\lambda \right) $ above, which is the
operation time for an ideal homogeneous coupling.

For a three-level flux qutrit, the transition frequency between two
neighboring levels can be varied from~5 GHz to~20 GHz. As an example, we
consider $\omega _{eg}/2\pi =7.5$ GHz and $\omega _{fg}/2\pi =12.5$ GHz, for
which we have $\Delta /2\pi =-2.5$ GHz. We set $\Delta _1/2\pi =0.75$ GHz
and $\Delta _2/2\pi =1.5$ GHz, which yields $\Delta _{a_1a_2}/2\pi =\Delta
_{b_1b_2}/2\pi =-0.75$ GHz, $\Delta _{a_1b_1}/2\pi =$ $\Delta _{a_2b_2}/2\pi
=-7.5$ GHz, $\Delta _{a_1b_2}/2\pi =-8.25$ GHz, and $\Delta _{a_2b_1}/2\pi
=-6.75$ GHz (Fig. 2). For simplicity, we choose $%
g_{a_jb_k}=g_{a_1a_2}=g_{b_1b_2}\equiv g_{cs}$ and $\Omega _{fe}=\Omega .$
Other parameters used in the numerical simulation are: (i) $\gamma _{\varphi
,e}^{-1}=2.5$ $\mu $s, $\gamma _{\varphi ,f}^{-1}=1.5$ $\mu $s, $\gamma
_{eg}^{-1}=5$ $\mu $s, $\gamma _{fe}^{-1}=2.5$ $\mu $s, $\gamma
_{fg}^{-1}=3.5$ $\mu $s (a conservative consideration, e.g., see Ref.~[12]);
and (ii) $\kappa _{a_j}^{-1}=\kappa _{b_j}^{-1}=10$ $\mu $s ($j=1,2$).

\begin{figure}[tbp]
\begin{center}
\includegraphics[bb=0 0 914 707, width=8.5 cm, clip]{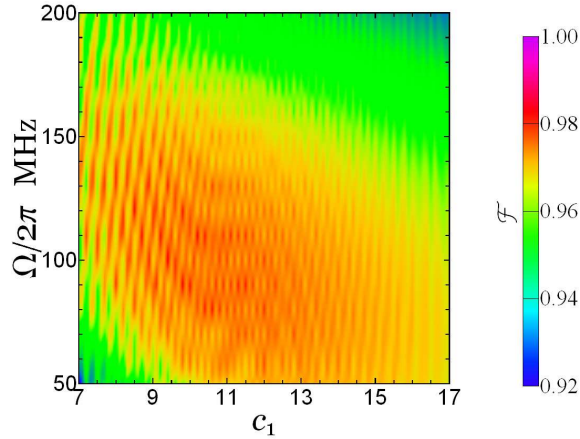} \vspace*{%
-0.08in}
\end{center}
\caption{(color online). Fidelity versus $c_1$ and $\Omega$. The figure was
plotted for $g_{cs}=0.4g_m$. Here, $c_1=\Delta_1/g_1$ is the normalized
detuning and $\Omega$ is the Rabi frequency.}
\label{fig:5}
\end{figure}

We now numerically calculate the fidelity for the two prepared EPR-pair
states. For given values of $\Delta _{1},\Delta _{2},$ and $g_{1},$ the
value of $g_{2}$ can be determined by Eq.~(15). We define $c_{1}=\Delta
_{1}/g_{1}$ and $c_{2}=\Delta _{2}/g_{2}.$ Based on Eq. (15), we have $c_{2}=%
\sqrt{\Delta _{2}/\Delta _{1}}c_{1}=\sqrt{2}c_{1}$ for the $\Delta _{1}$ and
$\Delta _{2}$ chosen above. To see how the inter-resonator crosstalk affects
the operation performance, in Fig. 4 we plot the fidelity versus $c_{1}$, by
choosing $\Omega /2\pi =100$ MHz and considering $%
g_{cs}=0,0.2g_{m},0.4g_{m},0.6g_{m},0.8g_{m},g_{m}.$ Here and below, $%
g_{m}=\max \{g_{1},g_{2},\mu _{1},\mu _{2}\}.$ From Fig. 4, one can see that
the effect of the inter-cavity crosstalk coupling is very small even when $%
g_{cs}=0.4g_{m}$, and a high fidelity $>97.86\%$ can be reached for $%
c_{1}=11 $ (corresponding to $c_{2}=11\sqrt{2}$)$.$ In this case, the
estimated operation time is $\sim 40$ ns. In the following analysis, we
choose $g_{cs}=0.4g_{m}.$ Note that according to the discussion in [44], a
smaller crosstalk $g_{cs}\leq 0.01g_{m}$ can be achieved with the typical
capacitive cavity-qutrit coupling illustrated in Fig. 1.

We now consider the dependence of the operation performance on the value of
the Rabi frequency $\Omega $ of the pulse. Figure~5 shows the fidelity
versus $c_{1}$ and $\Omega .$ From Fig.~5, one can see that the operation
performance strongly depends on the pulse Rabi frequency $\Omega $. On the
other hand, Fig.~5 shows that for $c_{1}\in \lbrack 7,17]$ ($c_{2}\in
\lbrack 7\sqrt{2},17\sqrt{2}]$)$,$ a high fidelity $\mathcal{F}\geq 92.4\%$
can be reached for a wide range of $\Omega $: $\Omega /2\pi \in \lbrack
50,200]$ MHz. Note that a pulse Rabi frequency $\Omega /2\pi \sim 300$ MHz
or higher was reported in experiments [76,77]. In Figure 5, the optimal
point is $c_{1}=10.2$ ($c_{2}=10.2\sqrt{2}$) and $\Omega /2\pi =110$ MHz,
for which the maximum fidelity of the joint state of the two prepared EPR
pairs is $\mathcal{F}_{\max }=98.42\%$, corresponding to the fidelities $%
\mathcal{F}_{a_{1}b_{1}}=99.05\%$ and $\mathcal{F}_{a_{2}b_{2}}=99.07\%$ for
the qubit pairs ($a_{1},b_{1}$) and ($a_{2},b_{2}$), respectively.

For $c_{1}\in \lbrack 7,17]$ and $c_{2}\in \lbrack 7\sqrt{2},17\sqrt{2}],$
we have $g_{1}/2\pi \in \lbrack 107,44]$ MHz, $g_{2}/2\pi \in \lbrack
151,62] $ MHz, $\mu _{1}/2\pi \in \lbrack 102,42]$ MHz, and $\mu _{2}/2\pi
\in \lbrack 143,59]$ MHz. The coupling strengths of these values are readily
achievable in experiments because a coupling strength $\sim 636$ MHz has
been reported for a superconducting flux device coupled to a one-dimensional
transmission line resonator [78]. For the transition frequencies of the
qutrit and the detunings given above, we have $\omega _{a_{1}}/\left( 2\pi
\right) \sim 11.75$ GHz, $\omega _{a_{2}}/\left( 2\pi \right) \sim 11$ GHz, $%
\omega _{b_{1}}/\left( 2\pi \right) \sim 4.25$ GHz, and $\omega
_{b_{2}}/\left( 2\pi \right) \sim 3.5$ GHz. Thus, for the values of $\kappa
_{a_{j}}^{-1}$ and $\kappa _{b_{j}}^{-1}$ used in the numerical simulation,
the required quality factors for the four resonators are $Q_{a_{1}}\sim
7.4\times 10^{5},$ $Q_{a_{2}}\sim 6.8\times 10^{5},$ $Q_{b_{1}}\sim
2.7\times 10^{5},$ and $Q_{b_{2}}\sim 2.2\times 10^{5},$ available in
experiments [21-23]. The analysis here demonstrates that the high-fidelity
generation of two EPR pairs of photonic qubits distributed in the four
resonators is feasible within present-day circuit QED techniques.

The prepared EPR pairs of photonic qubits can be read out by employing the
conventional approach [79], i.e., mapping the states of the photonic qubits
to superconducting qubits, whose states can be detected fast and accurately
[80]. One can also use an alternative method introduced in [69] to measure
the photonic qubits with a relatively fast speed and minimal action of
decoherence.

\begin{center}
\textbf{V. CONCLUSION}
\end{center}

We have presented an efficient method for simultaneously coupling multiple
pairs of resonators by using a qutrit as a coupler. This proposal
significantly reduces the effects of unwanted inter-resonator crosstalks
which are inherent in a circuit consisting of two or more resonators. We
showed that, under frequency matching conditions, the dynamics of the
resonators is described by an effective Hamiltonian, which can be used for
one-step generation of multiple EPR pairs of photonic qubits. Further, our
numerical simulation demonstrated that the obtained EPR states can have high
fidelities using present-day circuit QED technology. Finally, we note that
this effective Hamiltonian has other applications. For instance, it can be
directly applied to implement various quantum operations, such as the
simultaneous transfer or exchange of multi-photon quantum states between
spatially-separated resonators or cavities.

\begin{center}
\textbf{ACKNOWLEDGEMENTS}
\end{center}

This work was partly supported by the RIKEN iTHES Project, the MURI Center
for Dynamic Magneto-Optics via the AFOSR award number FA9550-14-1-0040, and
a Grant-in-Aid for Scientific Research (A). It was also partially supported
by the Major State Basic Research Development Program of China under Grant
No.~2012CB921601, the National Natural Science Foundation of China under
Grant Nos.~[11074062,~11374083], the Zhejiang Natural Science Foundation
under Grant No.~LZ13A040002, the funds of Hangzhou Normal University under
Grant Nos.~[HSQK0081,~PD13002004], and the funds of Hangzhou City for
supporting the Hangzhou-City Quantum Information and Quantum Optics
Innovation Research Team.

\end{document}